\documentclass[fleqn,usenatbib]{mnras}
\usepackage{graphicx,color} % Required for inserting images
\usepackage{newtxtext,newtxmath}
\usepackage[T1]{fontenc}
\usepackage{amsmath}

\usepackage{changes}

\DeclareRobustCommand{\VAN}[3]{#2}
\let\VANthebibliography\thebibliography
\def\thebibliography{\DeclareRobustCommand{\VAN}[3]{##3}\VANthebibliography}

\newcommand{\kms}{km\,s$^{-1}$}

\title[Transition from decaying to decayless kink oscillations]{Transition from decaying to decayless kink oscillations
of solar coronal loops}

\author[V.~M. Nakariakov et al.]{
Valery M. Nakariakov$^{1} \thanks{E-mail: V.M.Nakariakov@warwick.ac.uk}$,
Yu Zhong$^{1}$, Dmitrii~Y.~Kolotkov$^{1,2}$ %\thanks{MC and CS participated in this research as final-year MMathPhys students in 2021--2022},$
\\
$^{1}$Centre for Fusion, Space and Astrophysics, Physics Department, University of Warwick, Coventry CV4 7AL, UK\\
$^{2}$Engineering Research Institute \lq\lq Ventspils International Radio Astronomy Centre (VIRAC)\rq\rq, Ventspils University of Applied Sciences, Ventspils, LV-3601, Latvia
}

\date{Accepted 2024 June 09. Received 2024 June 07; in original form 2024 May 07}

\pubyear{2024}

\begin{document}
\label{firstpage}
\pagerange{\pageref{firstpage}--\pageref{lastpage}}
\maketitle

\begin{abstract}
The transition of an impulsively excited kink oscillation of a solar coronal loop to an oscillation with a stationary amplitude, i.e., the damping pattern, is determined using the low-dimensional self-oscillation model.  In the model, the decayless kink oscillations are sustained by the interaction of the oscillating loop with an external quasi-steady flow. 
The analytical solution is based on the assumption that the combined effect of the effective dissipation, for example, by resonant absorption, and interaction with an external flow, is weak. The effect is characterised by a dimensionless coupling parameter.
The damping pattern is found to depend upon the initial amplitude and the coupling parameter. The approximate expression shows a good agreement with a numerical solution of the self-oscillation equation. The plausibility of the established damping pattern is demonstrated by an observational example. Notably, the damping pattern is not exponential, and the characteristic decay time is different from the time determined by the traditionally used exponential damping fit. Implications of this finding for seismology of the solar coronal plasmas are discussed. In particular, it is suggested that a very rapid, in less than the oscillation period, decay of the oscillation to the stationary level, achieved for larger values of the coupling parameter, can explain the relative rareness of the kink oscillation events. 
\end{abstract}

\begin{keywords}
Sun: corona -- Sun: oscillations -- MHD -- waves
\end{keywords}

\section{Introduction}

The observational study and theoretical modelling of magnetohydrodynamic (MHD) wave processes in the corona of the Sun remain one of the key topics of modern solar physics \citep[see, e.g.,][]{2020ARA&A..58..441N}. Coronal MHD waves and oscillations attract attention mainly as possible agents which transfer the energy from lower to higher layers of the solar atmosphere \citep[e.g.,][]{2012RSPTA.370.3217P, 2020SSRv..216..140V}, and as natural probes of coronal plasma structures \citep[e.g.,][]{2024RvMPP...8...19N}. In addition, coronal MHD waves may play some role in solar flares \citep[e.g.,][]{2008ApJ...675.1645F, 2010PPCF...52l4009N, 2018SSRv..214...45M}.

One of the intensively studied coronal MHD wave phenomena are kink oscillations of coronal plasma loops \citep[e.g.,][for a recent comprehensive review]{2021SSRv..217...73N}. Typically, kink oscillations are detected either as transverse displacements of the loop in the plane of the sky with imaging instruments \citep[e.g.,][]{1999ApJ...520..880A}, or as periodic displacements of coronal emission lines in the spectral domain \citep[e.g.,][]{2012ApJ...759..144T}. 

In observations, kink oscillations appear in two regimes. The large-amplitude rapidly-decaying regime is characterised by the displacement amplitude exceeding several minor radii of the oscillating loop, and very short damping time, of a few oscillation periods. The oscillation period scales linearly with the oscillating loop length \citep[e.g.,][]{2016A&A...585A.137G, 2019ApJS..241...31N}. The decay time has been found to show a linear dependence on the oscillation period \citep[e.g.,][]{2002ApJ...576L.153O, 2016A&A...585A.137G, 2019ApJS..241...31N}. The quality factor defined as the ratio of the damping time to the oscillation period, decreases with the relative displacement amplitude to the minus two thirds \citep{2016A&A...590L...5G, 2020ARA&A..58..441N}. The oscillations are impulsively excited by the displacement of the loop from the equilibrium by a low coronal eruption \citep{2015A&A...577A...4Z}, or are associated with a flare \citep{1999ApJ...520..880A}. 

The decayless regime is characterised by low displacement amplitude, typically smaller than the minor radius of the oscillating loop \citep[e.g.,][]{2012ApJ...751L..27W, 2012ApJ...759..144T, 2022ApJ...930...55G, 2022RAA....22k5012G, 2022A&A...666L...2M, 2022AcASn..63....1G, 2023NatSR..1312963Z, 2024A&A...681L...4G}. The detection of the same decayless kink oscillations with two different EUV imagers confirmed that it is a natural phenomenon and not an instrumental artefact \citep{2022MNRAS.516.5989Z}.
Decayless kink oscillations may last for several tens of oscillation cycles with irregular variation of the oscillation parameters around a mean value \citep{2022MNRAS.513.1834Z}. The distribution of the detected oscillation amplitudes with the oscillation period is rather flat, indicating the lack of a resonant driver \citep{2016A&A...591L...5N}. 
The simultaneous detection of a decayless kink oscillation from two lines-of-sights separated by about 104$^{\circ}$, revealed a horizontal or weakly oblique linear polarisation of the oscillation \citep{2023NatCo..14.5298Z}. 

As in the decaying regime, the oscillation period scales linearly with the loop length \citep[e.g.,][]{2013A&A...560A.107A, 2023ApJ...944....8L}. The ratio of double the loop lengths and the periods corresponds to the phase speed consistent with the coronal kink speed of about a thousand \kms. Phase speeds about this value appear in decayless kink oscillations of both very long, $\sim$750~Mm, and very short, $\sim$5~Mm coronal loops \citep[see][respectively]{2023NatSR..1312963Z, 2023ApJ...946...36P}.
On the other hand, observations of decayless kink oscillations in short 5--20~Mm loops, also called coronal bright points, gave much longer periods, of several minutes, which showed no systematic scaling with the loop length \citep{2022ApJ...930...55G}. Similar oscillations have been observed in cold, transition region loops \citep{2024A&A...681L...4G}.  

Theoretically, kink oscillations are modelled as fast magnetoacoustic waves guided by perpendicular non-uniformity of equilibrium plasma parameters, such as the plasma density, of the $m=1$ symmetry, where $m$ is the azimuthal mode number \citep[e.g.,][]{1982SoPh...76..239E, 1983SoPh...88..179E}. An alternative interpretation has been suggested, in terms of an incompressive surface Alfv\'en wave \citep{2012ApJ...753..111G}. These two approaches converge in the infinite wavelength limit, while differ for finite wavelengths. The further discussion of this matter is beyond the scope of this paper. 

The rapid damping of decaying kink oscillations is traditionally associated with the effect of resonant absorption, based on the linear transformation of the collective fast magnetoacoustic mode into highly-localised incompressive Alfv\'enic motions in the vicinity of a narrow resonant layer where the local Alfv\'en speed coincides with the kink speed \citep[e.g.,][]{1992SoPh..138..233G, 2002A&A...394L..39G, 2002ApJ...577..475R, 2006RSPTA.364..433G}. The plausibility of kink oscillation damping by resonant absorption does not require the oscillating loop to have a circular cross-section or be embedded in a uniform environment \citep[e.g.,][]{2011ApJ...731...73P, 2024A&A...684A.154S}. This mechanism prescribes an exponential decay of the kink oscillation, with the damping time proportional linearly to the oscillation period. Furthermore, the initial stage of the kink oscillation damping is described by a Gaussian profile, resulting in a generalised, Gaussian-exponential damping pattern \citep[e.g.,][]{2012A&A...539A..37P, 2013A&A...551A..39H, 2019FrASS...6...22P}. Steep azimuthal shear flows generated by resonant absorption of kink oscillations have been shown to induce the Kelvin--Helmholtz instability (KHI) and uniturbulence which further enhance the damping \citep[e.g.,][]{2016A&A...595A..81M, 2021ApJ...910...58V}. This finding is in qualitative agreement  with the empirically determined dependence of the kink oscillation quality factor on the initial amplitude \citep{2016A&A...590L...5G, 2019ApJS..241...31N, 2021ApJ...915L..25A}. 

A mechanism responsible for sustaining decayless kink oscillations, i.e., compensating the energy decay, is subject to an intensive ongoing study. 
Decayless kink oscillations have been found in global simulations of the solar atmosphere from the convection zone to the solar corona \citep{2021A&A...647A..81K}.
\citet{2016ApJ...830L..22A, 2019ApJ...870...55G} suggested that a decayless oscillatory pattern can result from the combination of periodic brightenings produced by the KHI and the coherent motion of the KHI vortices, affected by finite spatial resolution and narrowband temperature response function. \citet{2023A&A...680L..15L} detected a decayless kink oscillation that lasted five oscillation cycles and occurred simultaneously with a quasi-periodic pulsation pattern in a nearby flare, suggesting a possible relationship between these two phenomena. 
An apparent decayless phase has been noticed in numerical simulations of the evolution of an impulsively excited kink oscillation in terms of the ideal incompressible model \citep{2015ApJ...803...43S}. A similar behaviour has been detected \citep{2024A&A...684A.154S} in the compressible model too. The decayless phase appears because of the interference of the torsional perturbations excited by resonant absorption, and hence can be considered as an \lq\lq extended\rq\rq\ phase of an impulsively excited kink oscillation. 
However, those mechanisms have difficulties explaining the long-durational appearance of decayless kink oscillations, as well as their regular appearance without any nearby flares or other impulsive energy releases. 
\citet{2024MNRAS.527.5741L} associated 5-min decayless kink oscillations observed in short loops with slow magnetoacoustic oscillations driven by photospheric oscillations. However, it is questionable whether high values of the plasma $\beta$, 0.3--0.5, required by this mechanism are consistent with those in coronal loops. Additionally, this mechanism can produce kink oscillations polarised solely within the plane of the loop. 
Random motions of footpoints of the oscillating loop can supply the energy required to sustain the oscillations too \citep{2020A&A...633L...8A}. The plausibility of this mechanism has been demonstrated analytically \citep{2021MNRAS.501.3017R, 2021SoPh..296..124R} and also through full MHD 3D numerical simulations \citep{2024A&A...681L...6K}. However, it is not clear whether this scenario can reproduce linearly polarised oscillations observed in the corona. Long-period decayless kink oscillations of short loops can be non-resonantly driven by photospheric and chromospheric oscillatory processes \citep[e.g.,][]{2023ApJ...955...73G}, but this phenomenon is out of scope of this paper. 

Quasi-stationary flows, with the evolution times much longer than the kink oscillation period, can sustain decayless kink oscillations through the nonlinear self-oscillatory mechanism. These flows could correspond to the most powerful, low-frequency part of the red noise spectrum of solar atmospheric movements \citep[e.g.,][]{2014A&A...563A...8A, 2014A&A...568A..96G, 2015ApJ...798....1I, 2016A&A...592A.153K, 2023FrASS..1099346A}. In this scenario, the energy to compensate damping losses is supplied by the negative friction between the oscillating loop and an external quasi-stationary flow. 
This process is similar to the vibration of a violin string, sustained by a moving bow. The concept of the self-oscillatory mechanism was demonstrated in a low-dimensional model by \citet{2016A&A...591L...5N}, and confirmed by full MHD 3D numerical simulations by \citet{2020ApJ...897L..35K}. Self-oscillation periods are determined by the oscillating loop length and the kink speed, and are weakly sensitive to random time variations of the model parameters and noisy driving \citep{2022MNRAS.516.5227N}. Furthermore, the amplitude experiences gradual fluctuations consistent with the observed behaviour \citep{2022MNRAS.513.1834Z}.

The transition from the decaying to decayless regimes \citep[see][for an observational example]{2013A&A...552A..57N} can provide empirical information essential for uncovering the mechanism behind decayless kink oscillations. In particular, \citet{2023Univ....9...95N} demonstrated that one can distinguish between the randomly driving and self-oscillatory scenarios by the pattern of the oscillation decay to the stationary, i.e., decayless amplitude. In the random-driving scenario, the damping is exponential, provided the damping mechanism is linear. In self-oscillations, the damping pattern was found to be super-exponential. The Markov chain Monte Carlo Bayesian comparison of the exponential, Gaussian--exponential, and super-exponential damping models, performed on ten kink oscillation events, showed the preference of the super-exponential damping \citep{2023MNRAS.525.5033Z}. However, the super-exponential damping is just a guessed pattern, and there is a need for the derivation of a damping pattern determined by parameters of the self-oscillator. 

The aim of this paper is to determine the functional expression for the evolution of the kink oscillation amplitude during the transition from the decaying to decayless regimes with the use of the low-dimensional self-oscillator model proposed in \citet{2016A&A...591L...5N}. In Section~\ref{sec:prob} we formulate the problem. In Section~\ref{sec:as} we obtain and study an asymptotic solution, and demonstrate its plausibility by an observational example in Section~\ref{sec:ao}. The obtained results are discussed, and conclusions are given in Section~\ref{sec:disc}.

\section{Problem formulation and governing equations}
\label{sec:prob}

Following the formalism developed in \citet{2016A&A...591L...5N, 2023Univ....9...95N}, we describe an impulsively excited kink oscillation using the low-dimensional self-oscillator model. The model is based on the assumption that the oscillation occurs in a kink resonator with the resonant frequency $\Omega_\mathrm{k} = \pi C_\mathrm{k}/L$ determined by the loop length $L$ and the kink speed $C_\mathrm{k}$. In this study we consider the loop to behave as a single-mode kink resonator. The oscillation is subject to a linear damping $\delta$ which represents, for example, the energy flow to Alfv\'enic motions by resonant absorption. Furthermore, the damping is compensated by the interaction of the loop with an external quasi-steady flow with the speed $v_0$ via negative friction. Thus, the governing equation is 
\begin{equation}
\label{goveEq1}
\ddot{\xi} +  \delta \dot{\xi} + \Omega_\mathrm{k}^2 \xi = F(v_0 - \dot{\xi}), 
\end{equation}
where $\xi$ is the displacement of the loop, the independent variable $t$ is time, and the right hand side term $F$ describes the interaction with the external flow.   
Taylor expanding the function $F$ and accounting for the first two terms, we obtain the self-oscillator ordinary differential equation,
\begin{equation}
\label{goveEq2}
\ddot{\xi} +  \left[ (\delta-\delta_\mathrm{v}) + \alpha \left(\dot{\xi}\right)^2\right]\dot{\xi} + \Omega_\mathrm{k}^2 \xi = 0, 
\end{equation}
where $\delta_\mathrm{v}$ and $\alpha$ are the constant coefficients of the two lowest order terms in the expansion \citep[e.g.,][]{2013PhR...525..167J}. 
For positive  $\Delta = \delta_\mathrm{v}-\delta$ and $\alpha$, Eq.~(\ref{goveEq2}) has a stable limit cycle solution 
\begin{equation}
\label{infamp}
\xi_\infty = \sqrt{4 \Delta/(3\alpha \Omega_\mathrm{k}^2)},
\end{equation} 
which corresponds to the decayless oscillatory pattern. 

With the use of the dimensionless variables,
\begin{equation}
\label{norm}
\displaystyle
Y = \left( \frac{3\alpha \Omega_\mathrm{k}^2}{\Delta} \right)^\frac{1}{2} \xi,\mbox{\ \ \ } \overline{t} = \Omega_\mathrm{k} t,
\end{equation}
Eq.~(\ref{goveEq2}) becomes
\begin{equation}
\label{goveEq3}
    Y''  + Y = \mu \left[ Y' - (Y')^3/3 \right],
\end{equation}
where $\mu = \Delta / \Omega_\mathrm{k}$ and the prime denotes the derivative with respect to the dimensionless variable $\overline{t}$. In the following the bar over the time variable will be omitted. The coefficient $\mu$ which accounts for both the dissipation and negative friction can be called the coupling parameter. 

Eq.~(\ref{goveEq3}) needs to be supplemented by initial conditions, for example,
\begin{equation}
\label{goveEq4}
    Y(0) = 2a, \mbox{\ \ \ }Y'(0) = 0,
\end{equation}
which correspond to the excitation of a kink oscillation by a displacement of the loop from an equilibrium \citep{2015A&A...577A...4Z}.

\section{Asymptotic solution}
\label{sec:as}

Considering the coefficient $\mu$ as a small parameter, we can determine an approximate solution to Eq.~(\ref{goveEq3}) using the standard asymptotic multi-scale method \citep[e.g.,][]{alma9924474017602466}. 
Let us look for a solution in a form of the asymptotic expansion by $\mu$,
\begin{equation}
\label{ass1}
Y(t) = Y_0(t, \tau) + \mu Y_1(t, \tau) + ...,
\end{equation}
where $\tau = \mu t$ is the \lq\lq slow\rq\rq\ time. 
Substituting expansion (\ref{ass1}) to Eq.~(\ref{goveEq3}), and combining terms in front of different powers of the small parameter $\mu$, we obtain in the lowest order, $\mu^0$,
\begin{equation}
\label{ass2}
Y_0'' + Y_0 = 0.
\end{equation}
Thus $Y_0 = A(\tau) \exp(it) + A^*(\tau) \exp(-it)$, where the superscript \lq\lq *\rq\rq\ denotes the complex conjugate. 

The terms of the first order, $\mu^1$, give us
\begin{eqnarray}
\label{ass3}
Y_1'' + Y_1 = 2 \left(- i \frac{d A}{d\tau} e^{it} + i  \frac{d A^*}{d\tau} e^{-it} \right) 
+ i A e^{it} - i A^* e^{-it}\\
-\frac{1}{3} \left( -i A^3 e^{3it} + i A^{*3} e^{-3it} + 3i|A|^2 A e^{it} - 3i|A|^2 A^* e^{-it} \right).\nonumber
\end{eqnarray}
The secular growth of the variable $Y_1$ requires the right hand side of Eq.~(\ref{ass3}) to be out of resonance with the natural frequency of the left hand side, thus 
\begin{equation}
\label{ass4}
\displaystyle
    \left\{ 
     \begin{array}{ll} \displaystyle
            2 i \frac{d A^*}{d\tau} - i A^* + i |A|^2 A^* = 0, \\
              \displaystyle   - 2 i \frac{d A}{d\tau} + i A - i |A|^2 A = 0 .
                \end{array}
    \right.
\end{equation}
Introducing $A(\tau) = R(\tau) \exp[i \theta (\tau)]$, where both the modulus $R$ and phase $\theta$ are real functions, 
we equate both imaginary and real parts to zero. Thus, we obtain from (\ref{ass4}) the separable ordinary differential equation 
\begin{equation}
\label{ass5}
2 \frac{d R}{d\tau} = R - R^3, 
\end{equation}
with the solution
\begin{equation}
\label{ass6}
\displaystyle
R(\tau) = \frac{R(0)} {\sqrt{e^{-\tau} + R^2(0) (1 - e^{-\tau})}}, 
\end{equation}
and a constant $\theta = \theta_0$. Applying initial conditions (\ref{goveEq4}), we get $ \theta_0 = 0$ and $R(0) = a$. 
Thus, the asymptotic solution to the initial value problem (\ref{goveEq3})--(\ref{goveEq4}) is
\begin{equation}
\label{ass7}
\displaystyle
Y(t) = 2R(\tau)\cos(t)-\frac{1}{3} \mu R^3(\tau)\sin^3(t) + ... .
\end{equation}
Thus, in the leading term only, we obtain 
\begin{equation}
\label{ass8a}
Y(t) \approx D(t) \cos(t). 
\end{equation}
where the factor in front of the cosine, $D(t)$, describes the evolution of the oscillation amplitude, 
\begin{equation}
\label{ass7a}
D(t) = \frac{2a} {\sqrt{e^{-t/\tau_\mathrm{as}} + a^2 (1 - e^{-t/\tau_\mathrm{as}})}}.
\end{equation}
Here, the quantity $\tau_\mathrm{as} = \mu^{-1}$ is the characteristic time of the amplitude evolution. In particular, it describes how rapidly the oscillation decays if the initial amplitude exceeds the stationary amplitude. 
In the following, we refer to the function describing the amplitude evolution, i.e., the envelopes of the oscillations, as the damping pattern. 

\begin{figure*}
   \centering
    \includegraphics[width=0.65\columnwidth]{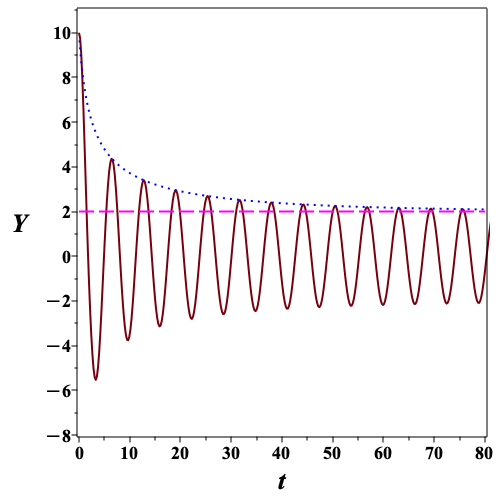}
     \includegraphics[width=0.65\columnwidth]{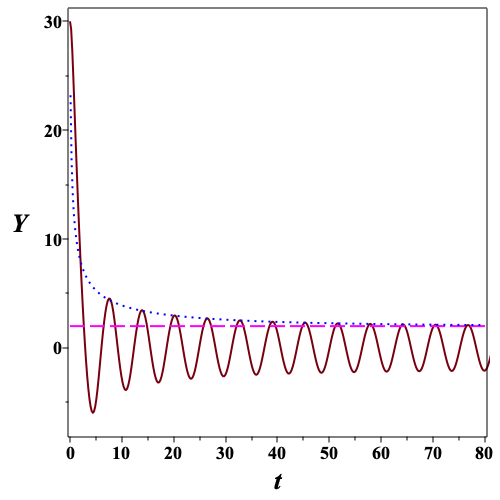}
      \includegraphics[width=0.65\columnwidth]{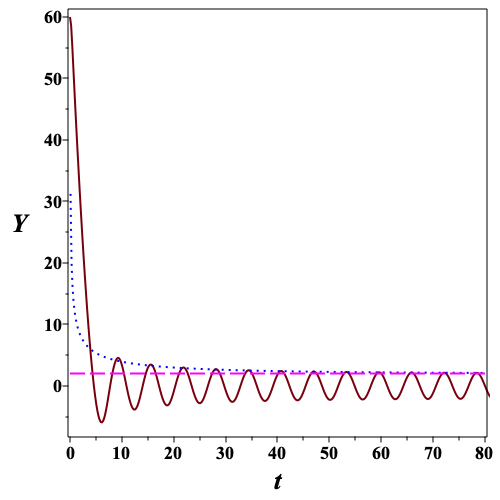}
    \caption{Comparison of the numerical and asymptotic solutions of the self-oscillator equation with the coupling parameter $\mu = 0.03$. The left, middle and right panels show the solutions with the initial amplitudes 5, 15 and 30 times higher than the stationary amplitude, respectively. Red solid curves show numerical solutions. Magenta dashed horizontal lines indicate the stationary (decayless) amplitude which is equal to 2 in all cases. Blue dotted curves show asymptotic solutions. The oscillation period is $2\pi$ time units.}
    \label{fig:1}
\end{figure*}

In the limit $t\to + \infty$, the amplitude approaches the value of 2, which is the stationary amplitude of the self-oscillation, i.e., the amplitude of the decayless regime. 
In the phase portrait of Eq.~(\ref{goveEq3}), this corresponds to a stable limit cycle \citep[e.g.,][]{2013PhR...525..167J}. 
In dimensional quantities, it is equivalent to the value $\xi_\infty$ given by Eq.~(\ref{infamp}). The oscillation either grows or decays to the stationary amplitude, depending upon whether the initial amplitude is higher or lower than the stationary amplitude, respectively, see Fig.~5 in \citet{2016A&A...591L...5N}. 
Fig.~\ref{fig:1} shows that for the initial amplitudes up to about an order of magnitude higher than the stationary amplitude, the approximate asymptotic damping pattern given by Eq.~(\ref{ass7a}) is in a good agreement with the numerical one. However, for a large initial amplitude, there is a visible discrepancy. 

\begin{figure}
   \centering
    \includegraphics[width=\columnwidth]{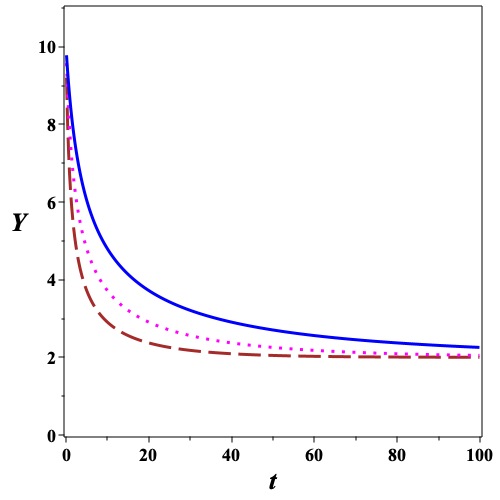}
    \caption{Damping patterns of a self-oscillation approaching the stationary amplitude for different values of the coupling parameter $\mu$: the blue solid curve corresponds to $\mu=0.015$, the magenta dotted curve to $\mu=0.03$, and the brown dashed curve to $\mu = 0.06$. The oscillation period is $2\pi$ time units. }
    \label{fig:2}
\end{figure}

In the dimensionless case described by Eq.~(\ref{goveEq3}), the amplitude of the stationary self-oscillation is independent of the initial amplitude $2a$ and the coupling parameter $\mu$. The value of $\mu$ controls the steepness of the damping pattern, as shown in Fig.~\ref{fig:2}, where the damping envelopes are calculated from Eq.~(\ref{ass7a}). Large values of $\mu$, i.e., the stronger coupling of the loop motion with the external quasi-steady flow, result in more rapid damping of the oscillation to the stationary amplitude. 

For large values of the initial amplitude, $a \gg 1$, i.e., when the initial amplitude is very much larger than the stationary amplitude, expression (\ref{ass7a}) reduces to
\begin{equation}
\label{ass8}
   D(t) \approx  2 \Big[ 1 - \exp(-t/\tau_\mathrm{as})\Big]^{-1/2},
\end{equation}
and becomes independent of $a$. 

\begin{figure*}
   \centering
    \includegraphics[width=\columnwidth]{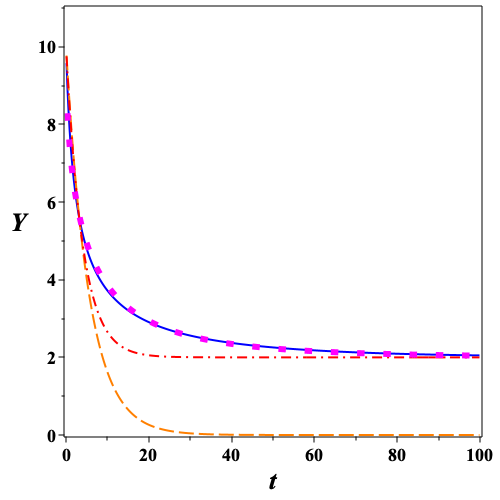}
        \includegraphics[width=\columnwidth]{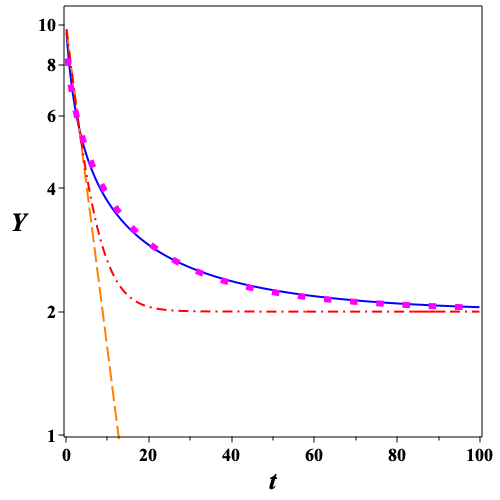}
    \caption{Comparison of different models of damping patterns of kink oscillations with the initial amplitude 10. The blue solid curve corresponds to the approximate asymptotic damping pattern of a self-oscillation for the coupling parameter $\mu = 0.03$.
    The magenta dotted curve shows a super-exponential profile with $\tau_\mathrm{se}=5$ and $p=0.55$.
    The orange dash-dotted curve shows the exponential damping with $\tau_\mathrm{ec}=4$ to the stationary amplitude. 
    The brown dashed curve is the exponential damping with $\tau_\mathrm{e}=5.5$. The left and right panels have linear and semi-logarithmic vertical axes, respectively.
    The oscillation period is $2\pi$ time units.}
    \label{fig:3}
\end{figure*}

Fig.~\ref{fig:3} shows damping patterns given by different models. 
Here we compare the asymptotic solution given by Eq.~(\ref{ass7a}), the exponential damping
\begin{equation}
\label{decexp}
    D_\mathrm{e}(t) = 2a \exp(-t/\tau_\mathrm{e}),
\end{equation}
which is prescribed by the resonant absorption of a free kink oscillation \citep[e.g.,][]{2002ApJ...577..475R};
the exponential damping to the oscillation with the constant amplitude $2$,
\begin{equation}
\label{deccon}
    D_\mathrm{ec}(t) = (2a-2) \exp(-t/\tau_\mathrm{ec}) +2,
\end{equation}
which described a decayless oscillation sustained by a random driver \citep{2023Univ....9...95N};
and the super-exponential damping to the oscillation with the constant amplitude,
\begin{equation}
\label{superexp}
    D_\mathrm{se}(t) = (2a-2) \exp[-(t/\tau_\mathrm{se})^p] +2,
\end{equation}
which was an empirical guess made in \citet{2023Univ....9...95N}.
In Eqs.~(\ref{decexp})--(\ref{superexp}), the constants $\tau_\mathrm{e}$, $\tau_\mathrm{ec}$, and $\tau_\mathrm{se}$ are the characteristic damping times, and $p$ is the super-exponential decay index. All patterns begin at the initial amplitude $2a$. 
Here, we do not use the Gaussian-exponential damping pattern, as the fitted oscillatory curve does not demonstrate this property. 
Interestingly, the super-exponential decay pattern is very similar  to the asymptotic pattern. The semi-logarithmic plot suggests the possibility to distinguish between the traditionally used exponential damping pattern which appears to be a straight line, and the other patterns. 

\begin{figure}
%   \centering
    \includegraphics[width=\columnwidth]{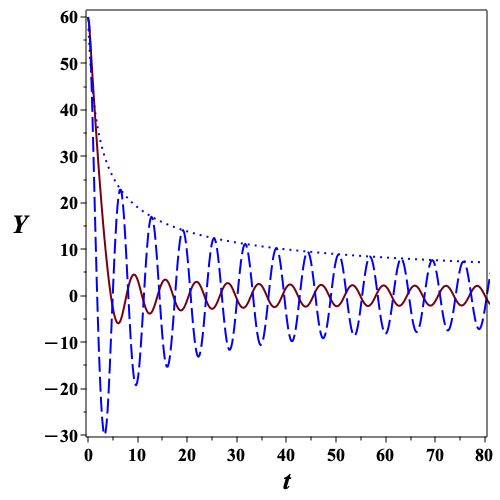}
    \caption{The effect of the coupling parameter $\mu$ on the damping pattern of a self-oscillation. The solid burgundy curve shows the numerical solution of the dimensionless self-oscillator equation in the case $\mu = 0.03$, the dashed blue curve shows the case $\mu = 10^{-3}$. The dotted blue curve demonstrates the asymptotic damping pattern for $\mu = 10^{-3}$. In all cases, $a=30$, and the stationary amplitude is 2. }
    \label{fig:4}
\end{figure}

Fig.~\ref{fig:1} shows that for a fixed value of the coupling coefficient $\mu$ the match of damping pattern (\ref{ass7a}) with the numerical solution of the self-oscillation equation worsens with the increase in the ratio of the initial and stationary amplitudes. However, for the same initial amplitude the transition to the decayless regime may be much slower. Fig.~\ref{fig:4} demonstrates that for smaller damping parameters the damping pattern is consistent with the asymptotic solution, as one would expect for an asymptotic solution.

\section{An observational example}
\label{sec:ao}

The damping patterns given by Eqs.~(\ref{ass7a}) and (\ref{ass8}) are different from the exponential damping patterns traditionally used in the analysis of decaying kink oscillations, in particular, in the catalogues \citep{2016A&A...585A.137G,2019ApJS..241...31N}. Furthermore, \citet{2023MNRAS.525.5033Z} demonstrated that non-exponential damping patterns may be more consistent with observations. 
Here, we illustrate the plausibility of the asymptotic damping pattern by reanalysing a kink oscillation event.

\begin{figure}
   \centering
    \includegraphics[width=\columnwidth, trim=0.5cm 0.1cm 1.6cm 0em, clip]{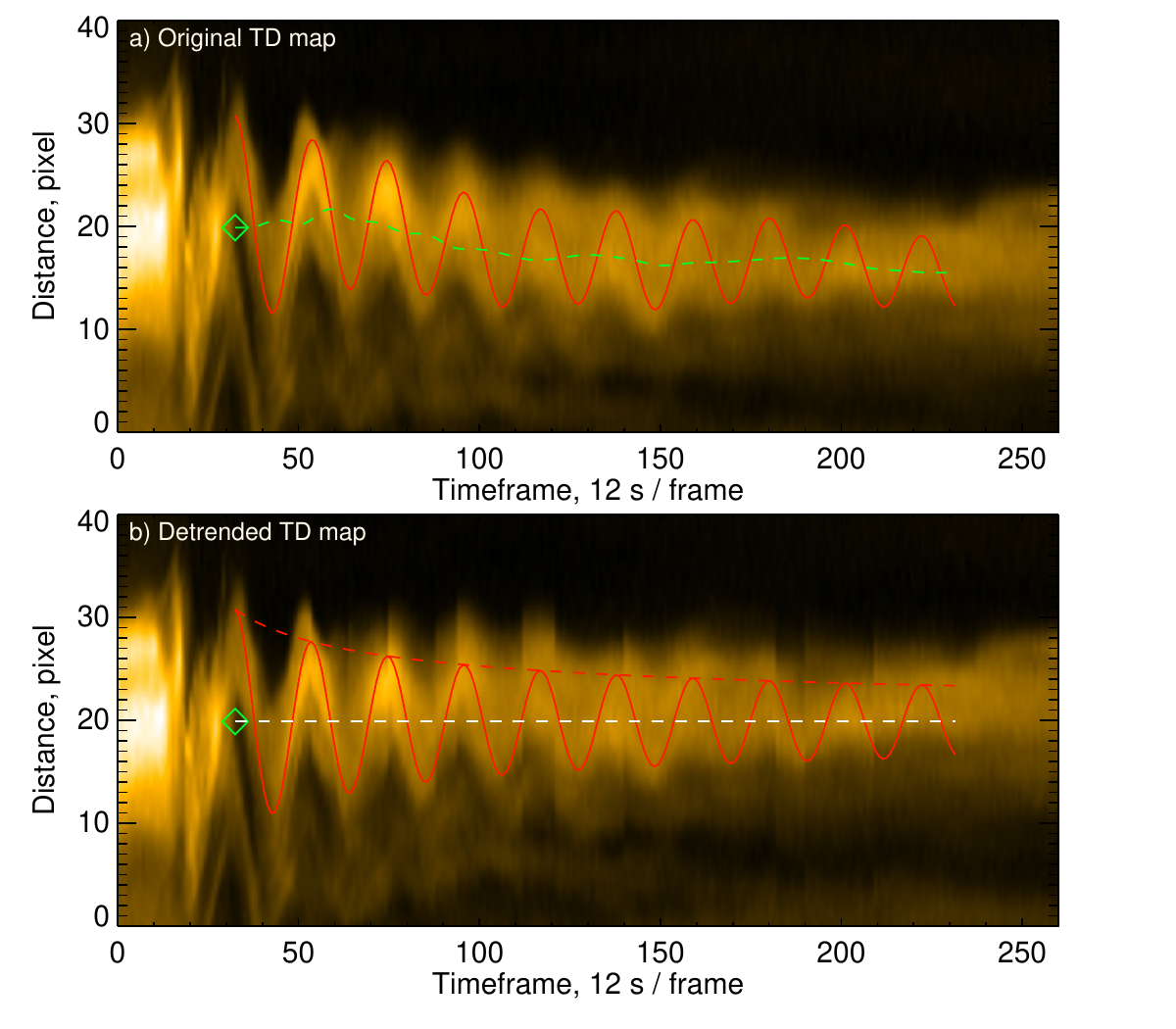}
    \caption{A kink oscillation event on May 30, 2012, in active region NOAA~11494, showing the transition from the decaying to decayless regimes. The red solid and dashed curves correspond to the oscillation pattern and its damping pattern, respectively. The green dashed curve represents the averaged background trend of the oscillatory signals. The time--distance map shown in panel (b) is obtained by subtracting the averaged trend from that in panel (a).}
    \label{fig:5}
\end{figure}

The event of our interest was observed by \citet{2013A&A...552A..57N} on May 30, 2012, in active region NOAA~11494, with the Atmospheric Imaging Assembly (AIA) on the Solar Dynamics Observatory spacecraft. The oscillation clearly showed the transition from the decaying to the decayless regimes. 
The time--distance map was made for a 5-pixel-wide slit $\{(-973.0^{\prime \prime}, -366.0^{\prime \prime}), (-995.5^{\prime \prime}, -330.0^{\prime \prime})\}$, perpendicular to the oscillating loop near the apex in the stack of 171\AA\ images in the time interval 08:57:42--09:49:42~UT, see Fig.~\ref{fig:5}a. 

The damping pattern described by Eq.~(\ref{ass7a}) was compared with the observed behaviour of the kink oscillation through the following steps.
First, we quantify the oscillating signals both at the loop centre and boundary. From the high-contrast upper loop boundary (see Fig.~\ref{fig:5}a), we extracted the boundary displacement $\xi_{1}$ manually, and repeated this operation three times to reduce uncertainty. In addition, the displacements were determined automatically by four loop tracking methods described in \citet{2023MNRAS.525.5033Z} to obtain signals for the loop center displacements $\xi_{2}$--$\xi_{5}$. 

Then, we determine the background trend of the oscillatory signals $\xi_{1}$--$\xi_{5}$ with the detrending method described in \citet{2022MNRAS.513.1834Z}. After identifying the oscillation crests and troughs, i.e., the extrema of the displacement positions, we used the spline interpolation to obtain the background trends $\rm{BG}_{1}$--$\rm{BG}_{5}$ of the signals $\xi_{1}$--$\xi_{5}$, respectively. In the following we use a detrended signal obtained by subtracting the average background trend from the average signal. 

Parameters $a$ and $\Omega_\mathrm{k}$ of the asymptotic damping pattern are determined from the observed signal. We chose the first oscillation crest at 09:04:12~UT as a starting point and extracted the initial amplitude $A_0 =2.90$~Mm. The decayless amplitude is estimated as $A_\infty=0.44$~Mm from the detrended signal. % \textbf{$\xi_{1}$ subtracted by $\rm{BG}_{1}$}. 
Then, the ratio of the initial and stationary amplitudes allows us to estimate the dimensionless parameter $a = A_\infty/A_0 \approx 6.6$. The average time between maxima (or minima) of the oscillation gives us the oscillation period and hence the angular frequency, $\Omega_\mathrm{k} \approx 1.50$~rad $\mathrm{s}^{-1}$. 

As the normalised scale in Eq.~(\ref{norm}) can be expressed as $\sqrt{\mu/(3\alpha \Omega_\mathrm{k})}$ and calculated from the ratio of $A_0$ to $2a$, we have the relation $\alpha = {4\mu a^2}/({3\Omega_k A_0^2})$. It leaves us with a single free parameter in the expression for the damping pattern, $\tau_\mathrm{as} = \mu^{-1}$. The best-fitting value of this parameter, $\mu=1.5\times 10^{-3}$, is then determined by the least square method. 

Fig.~\ref{fig:5} demonstrates a fairly good agreement between the asymptotic damping pattern described by Eq.~(\ref{ass7a}) with the observed transition of a kink oscillation to the decayless regime.

\section{Discussion and Conclusions}
\label{sec:disc}

In terms of the low-dimensional self-oscillation model, we consider the evolution of amplitude of a kink oscillation excited by an initial displacement of the loop from the equilibrium. Because of the interaction of the oscillating loop with an external quasi-steady flow with a time scale much longer than the oscillation period, the oscillation approaches a stationary amplitude regime which may be responsible for the decayless kink oscillations ubiquitously present in the corona. Assuming the combined effect of the dissipation and coupling with an external flow to be weak, we determined the asymptotic expression which describes the transition of a decaying kink oscillation to the stationary amplitude. This dependence differs from the exponential one. In dimensionless variables, when the time is measured in oscillation periods, and the amplitude is in units of the stationary amplitude, the damping pattern includes two free parameters: the initial amplitude and a parameter which accounts for the dissipation and coupling, see Eq.~(\ref{ass7a}). 
The dependence of the damping of kink oscillations on the initial amplitude is in a qualitative agreement with statistical results obtained by \citet{2021ApJ...915L..25A}. The damping time is determined by the difference between the negative friction between the oscillating loop and the external flow, and the effective damping due to, e.g., resonant absorption. The low-dimensional model employed here does not allow us to link this with specific physical quantities, which is a task of a future full-scale modelling. The derived damping pattern is found to be in a fairly good agreement with an observational example of the transition from the decaying to decayless regimes of kink oscillations \citep{2013A&A...552A..57N}. The comparison of different damping scenarios, and its use for revealing the mechanism for sustaining the decayless regime require a larger number of such events. 

If the initial amplitude exceeds significantly the stationary amplitude of the self-oscillation, the damping pattern becomes independent of the initial amplitude. The oscillation still asymptotically diminishes towards  the stationary value. In observations, the stationary amplitude may be below the threshold of detection, determined by the spatial resolution and sensitivity of the instrument. In this case, the kink self-oscillation appears as a regular decaying oscillation. However, the damping pattern is not exponential, and the characteristic decay time is different from the time determined by the exponential fit. In certain cases, the asymptotic solution shows a good fit with the super-exponential damping pattern proposed in \citet{2023Univ....9...95N}. It is consistent with  the preference for the super-exponential damping model over exponential and the Gaussian--exponential damping models, established by the analysis of ten randomly selected kink oscillation events \citep{2023MNRAS.525.5033Z}. If the prevalence of the non-exponential damping pattern is confirmed through the re-analysis of a statistically significant number of decaying kink oscillations, e.g., with the use of the catalogue of \citet{2019ApJS..241...31N}, it would suggest the necessity of recalculating of the oscillation damping times and their scaling with other oscillation parameters, such as periods and amplitudes. In turn, it would require the modification of the seismological techniques based on the use of the kink oscillation damping time \citep[e.g.,][]{2003ApJ...598.1375A, 2007A&A...463..333A, 2014RAA....14..805P, 2018ApJ...860...31P, 2018AdSpR..61..655A, 2019A&A...625A..35A, 2022FrASS...926947A}. Furthermore, the independence of the damping pattern of the initial amplitude, derived in the limit of a large initial amplitude does not contradict the results of \citet{2021ApJ...915L..25A}, as that study was based on the assumption that the initial amplitude is finite and the oscillation decays to a zero amplitude. In contrast, our study addresses the opposite limit, assuming that the oscillation decays to a finite amplitude. The combination of these two approaches is an interesting future topic. 

Another important question is whether the decay pattern is affected by the apparent decayless phase caused by the interference of the Alfv\'en continuum perturbations excited by resonant absorption, found in numerical simulations by \citet{2015ApJ...803...43S} and \citet{2024A&A...684A.154S}. It is clear that the kink perturbations will eventually decay toward zero from energetics considerations. However, this effect may modify the decay pattern.

For a broad range of the problem parameters, the oscillation period remains almost constant, which is consistent with previous theoretical and numerical findings \citep[e.g.,][]{2022MNRAS.516.5227N, 2020ApJ...897L..35K, 2023Univ....9...95N}. It has already been demonstrated that the kink oscillation period is insensitive to the specific mechanism for sustaining the decayless regime \citep{2022MNRAS.516.5227N}, and hence the seismological techniques based upon the oscillation period remain valid. The generalisation of the zero-dimensional self-oscillation model to a one-dimensional model \citep[cf.][]{2020A&A...633L...8A}, is necessary for the validation of seismological techniques based on the use of the ratio of various parallel harmonics of the kink oscillation \citep[e.g.,][]{2009SSRv..149....3A, 2013ApJ...767..169L}.

For large initial amplitudes and coupling parameters, the oscillation approaches the decayless amplitude in less than one oscillation period, see, e.g., the right panel of Fig.~\ref{fig:1}. This finding may explain the relative rareness of decaying kink oscillations of coronal loops.
If the decayless amplitude is below the detection threshold, the motion of the displaced loop appears to be a gradual return to the equilibrium, without its overshooting and oscillations around it.

The findings presented in this work are obtained in terms of a zero-dimensional model which neglects a number of effects. However, this approach allows one to explore potentially the most important features of the phenomenon of interest. The validation of our findings requires and motivates further observational and numerical studies.

\section*{Acknowledgements}
The data used are courtesy of the SDO/AIA team. VMN and DYK acknowledge support from the Latvian Council of Science Project No. lzp2022/1-0017. DYK acknowledges support from the STFC consolidated grant ST/X000915/1. 

\section*{Data Availability}
The observational data used in this study were obtained from the SDO/AIA instrument (\url{http://jsoc.stanford.edu/ajax/exportdata2.html?ds=aia.lev1/}) and are publicly available.

\bibliographystyle{mnras}
\bibliography{decay}

\bsp	% typesetting comment
\label{lastpage}
\end{document}